\documentclass[12pt,a4paper]{article}
\usepackage[english]{babel}
\usepackage{amsmath,amsthm}
\usepackage{amsfonts}
\usepackage{amssymb}
\usepackage{graphicx}

\title{Validation of the 3-under-2 principle of cell wall growth in Gram-positive bacteria by simulation of a simple coarse-grained model}
\author{M. Dinh$^1$, L. Strafella$^2$, P. Flores$^2$, A. Chastanet$^2$, \\ R. Carballido-L\'opez$^2$ and V. Fromion$^1$ \\[2ex] {\footnotesize $^1$ MaIAGE, INRA, Universit\'e Paris-Saclay, F-78350 Jouy-en-Josas, France} \\[-1ex] {\footnotesize $^2$ MICALIS, INRA, AgroParisTech, Universit\'e Paris-Saclay, F-78350 Jouy-en-Josas, France}}

\begin{document}

\maketitle

\begin{abstract}
The aim of this work is to propose a first coarse-grained model of {\it Bacillus subtilis} cell wall, handling explicitly the existence of multiple layers of peptidoglycans. In this first work, we aim at the validation of the recently proposed ``3-under-2'' principle.
\end{abstract}

\section{Introduction}
Bacterial cells are enclosed in the envelope, a multi-layer boundary, that confers shape and protection while allowing communication and exchanges with the environment. This casing is composed of a soft watertight lipid bilayer, the cytoplasmic membrane (plus an additional “outer membrane” in Gram-negative bacteria), and the cell-wall (CW), a rigid sugar-based exoskeleton composed of a peptidoglycan (PG) meshwork supporting linked teichoic acids (TA).

The bacterial CW is essential to most bacteria, the PG mesh conferring cellular shape and resistance to internal osmotic pressure as well as external stresses, and as such is the target of many antibiotics including Penicillins, Cephalosporins or Glycopeptides. It is suspected that it is a highly dynamic structure, constantly synthesized and remodeled as the cell cycle progresses and as the bacterium adapts to its environment, see {\it e.g.} \cite{Poo:76,Poo:76b,BuP:84,DCV:88,VoS:10,Hol:98,SKW:10}. Many enzymes involved in PG synthesis has been discovered along the years and decades of studies have revealed with great details the composition of the PG, its precursors and the chain of assembly leading to the building of this scaffold. To briefly summarize this series of events, the basic bricks, made of a disaccharide (N-acetylmuramic acid [MurNac] plus N-acetylglucosamine [GlcNac]) bearing a pentapeptide are synthesized in the cytosol, flipped through the membrane before being assembled into glycan chains by transglycosylases, and finally cross-linked to the existing mesh by transpeptidation of their side peptides (for a review, see \cite{Hol:98}). These reactions and their main catalytic enzymes have been known for decades but the resulting ultra-structure of the meshwork and the molecular mechanisms that control its morphogenesis are still largely unknown and remain a challenging problem in bacterial cell biology.

Recent technological improvements have brought new hints on the CW architecture and control of PG synthesis over the recent years. First, the use of electron cryo-tomography (ECT) have shed light on the organization of the PG in both Gram-negative (G-) and Gram-positive (G+) bacteria \cite{GCJ:08,BGRJ:13}. In {\it E. coli} and {\it Caulobacter crescentus}, both G-, ECT revealed thin, mostly circumferentially organized structures, compatible with a “layered” model of sacculus where PG strands would run parallel to the membrane \cite{GCJ:08}. These observations refuted the “scaffold” model \cite{MBHF+:06} in which PG strands would be inserted vertically, {\it i.e.} perpendicularly to the membrane. For G+ bacteria, the structure of the sacculus is harder to observe, due to the thickness of the material \cite{MaB:05}. Again, ECT, combined with molecular dynamics simulation, had a decisive weight in support of a circumferential organization of PG strands \cite{BGRJ:13}. Indeed, three models of PG structures have been proposed for G+ bacteria along the years: \emph{i)} the “scaffold” model \cite{DTE:05,MBHF+:06}, \emph{ii)} the “layered” model \cite{Ghu:68} and more recently \emph{iii)} the “coiled cable” model, based on atomic force microscopy images, where PG strands would form a 25 nm thick rope-like structure coiled in a 50 nm wide tube that itself would be coiled around the circumference of the cell \cite{HKHF:08}. This last model has yet to be backed up by additional results and, to the contrary, the observation with electron cryo-microscopy \cite{MaB:05} and more recently ECT \cite{BGRJ:13} are arguing against the existence of such coiled cables. In addition, this model postulates a very complex three-dimensional organization for which we lack both a mechanism and a required synthetic complex that could account for it. The thickness of the PG, the observations of curling fragments of purified PG and thickening of PG during relaxation, strengthen by molecular dynamics simulations, are all in favor of the layered model of PG \cite{BGRJ:13}.

Second, and in agreement with the above-mentioned observations, using advanced fluorescent microscopy techniques (total internal reflection fluorescence (TIRF), scanning confocal microscopy, and single particle tracking), several groups have followed live movements of essential proteins supposedly part of the so-called PG-elongation machinery (PGEM) \cite{ECBF+:11,GBWZ+11,vTWF+:11,BCYC+:17}. The analysis and quantification of the dynamics of these PGEM showed, in G+ and G- bacteria, directional circular motion perpendicular to the long axis of the cells suggesting an assembly and/or cross-linking of the PG strands alike.

Finally, recent advances have allowed the computational community to take a decisive step towards the development of coarse-grained models describing not only the structure and properties of the CW but also the dynamics of the components and enzymes during the cell cycle.\footnote{There is also  models describing the CW as a global and continuous elastic structure comparable to some extent to an elastic balloon with a non-uniform thickness, inflated by the osmotic pressure, see e.g. \cite{SCWJ+:10,MRGH:13}.} Such a coarse-grained model makes it possible to couple global aspects of the morphogenesis with more local molecular mechanisms related to the wall synthesis. So far,  models have  been developed only for G- CW such that of {\it E. coli} for which the thin structure (3-6 nm) is compatible with a mono-layered PG mesh \cite{VoS:10}. This simple organization allows to consider the CW as a 2D elastic network composed of strands of glycan, cross-linked by stretchable peptides \cite{HMWG+:08,FWH:11,NGBJ:15}. These modeling approaches have proven useful to test predictions on the fine mechanisms underlying the insertion of strands during growth such as the composition of the synthetic machineries, orientation of the enzymes, the sequence of enzymatic reactions, {\it etc} \cite{NGBJ:15}. Thus, there is today a solid line of experiments propped by mathematical modeling supporting a cut-and-insertion strategy of new PG strands in a monolayered CW in G- bacteria.

To the contrary, for G+ CW there is currently,  to our knowledge, no theoretical model describing the 3D structure and its assembly process, to the exception of our recently proposed 3-under-2 model \cite{BCYC+:17}. No coarse-grained mathematical model describing G+ PG have been described yet  most likely because of the more complex nature of its organization, composed of multiple layers of PG. Indeed, if evidences suggest that PG strands run roughly circumferentially and parallel to each other, they do not explain how they would be organized in a 3D structure, 30-50 nm thick, neither how the new material would be inserted and would mature, allowing the expansion of the sacculus during cell growth. While in {\it E. coli}, the elongation of a mono-layered PG can be simply explained through the insertion of new PG materials in the existing mesh (see {\it e.g.} \cite{BuP:84}), this mechanism does not support the establishment and maintenance of a multi-layered structure across generations. Instead, it is postulated that in G+ bacteria, new layers of PG are added to the innermost face of the CW, pushing outwards the previous layers in an “inside-to-outside” process; the outermost layers being eventually degraded, accounting for the observed PG turnover (see \cite{Poo:76,Poo:76b,DBKW:81,MKDS:84,DCV:88}). But this does not explain how the scaffold expand, what is the reticulation of the network, how it matures, how and where the PGEM works, how they control PG hydrolysis...

Recently, we proposed a theoretical synthesis model, the so called ``3-under-2'' principle, to explain how addition of new PG material to the existing mesh could account for expansion of the network while maintaining a constant thickness \cite{BCYC+:17}. In this model, the newly synthesized glycan strands are cross-linked to the innermost layer in a 2/1 geometry, doubling the number of strands at each layer, while insuring lateral expansion, thickness maintenance and cell integrity during growth. Here, the aim of this work is to, based on a reasonable set of assumptions, propose a first coarse-grained model describing a {\it B. subtilis} multi-layered sacculus, and put the so called ``3-under-2'' principle to the test. Using a minimalistic bi-layers CW, our results show that ...

\section{Consequence of the morphogenesis on the ultra-structure}

\begin{figure}[hbtp]
  \centering
  \includegraphics[width=7cm]{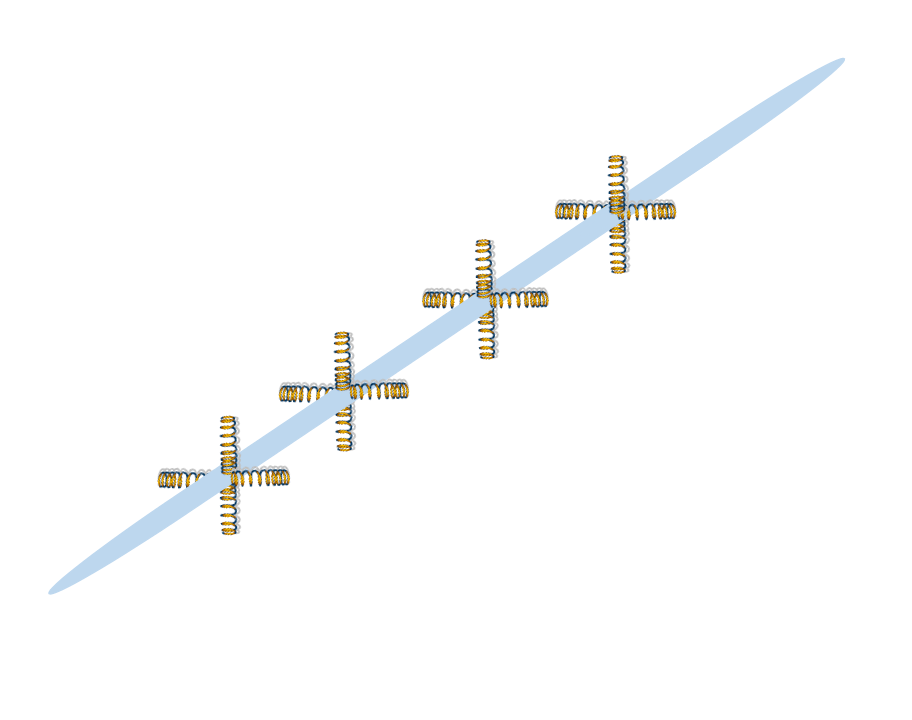}
  \caption{Assumption on a PG strand structure used for explanation}
  \label{fig:pgStrandAssume}
\end{figure}

For the sake of clarity, we first assume that a PG has 4 potential bonds as represented in Figure \ref{fig:pgStrandAssume}.
The CW ultra-structure is obtained by the superposition of several layers, with the same structure as the one used in \cite{HMWG+:08,NGBJ:15}. Consequently,  it is a planar model with a 90 degree angle between two successive peptidic cross-links. For sake of simplicity, we assume that the CW ultra-structure is regular.

\begin{figure}[hbtp]
  \centering
  \includegraphics[width=13cm]{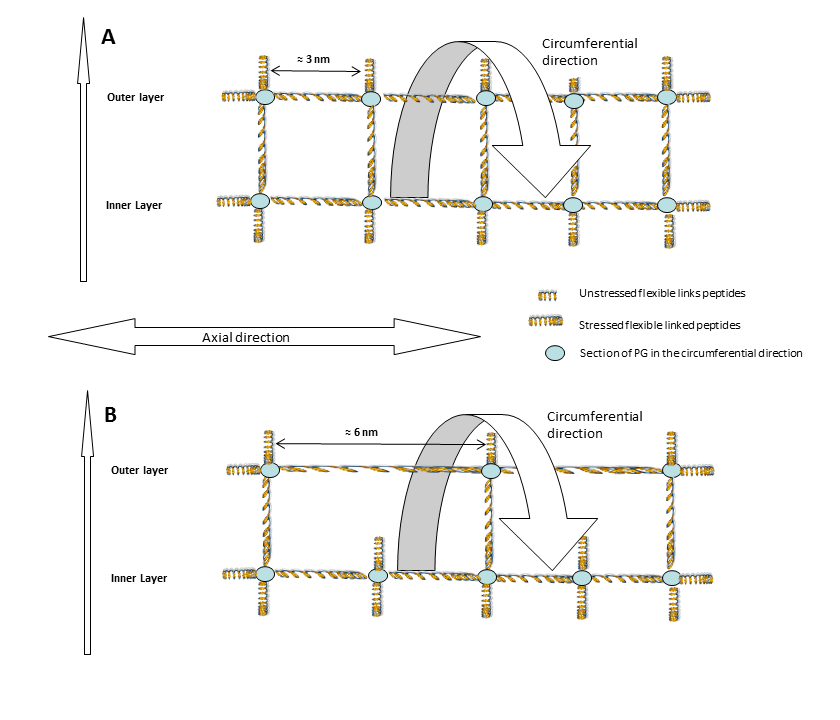}
  \caption{(\textbf{A}) A potential bi-layers ultra-structure. (\textbf{B}) A bi-layers CW ultra-structure with the ``3-under-2'' morphogenesis principle.}
  \label{fig:structureAssume}
\end{figure}

\begin{figure}[hbtp]
  \centering
  \includegraphics[width=13cm]{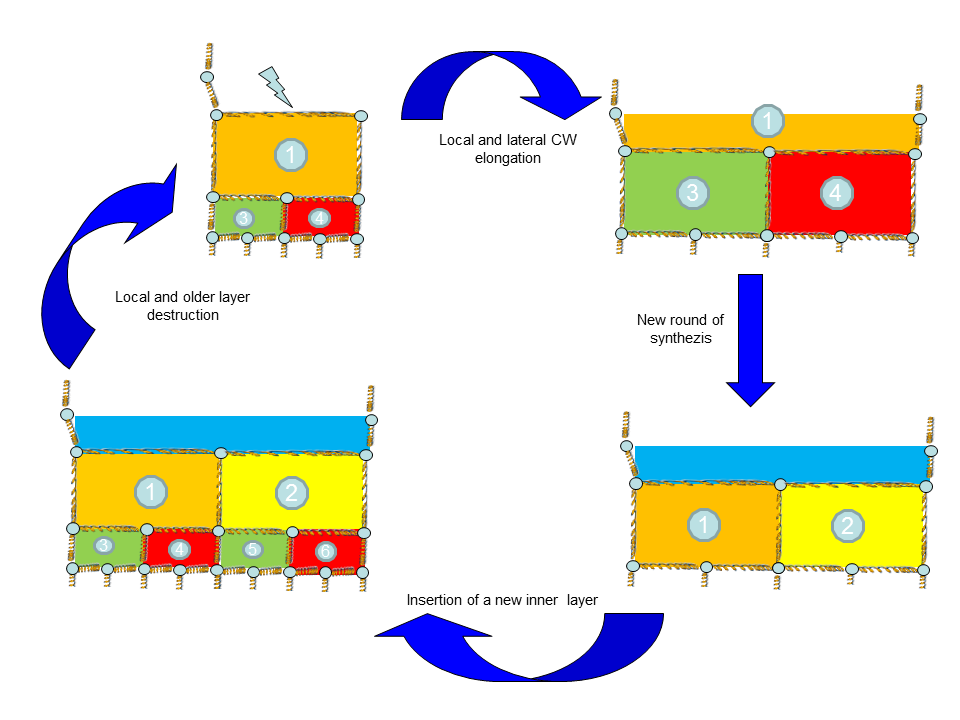}
  \caption{Insertion cycle of PG strands with the ``3-under-2'' morphogenesis principle.}
  \label{fig:insertionCycleAssume}
\end{figure}

With the above assumptions on the PG strand, it is possible to obtain different configurations for
a bi-layers CW as illustrated in Figure \ref{fig:structureAssume}.
In order to ensure that a PG layer insertion leads to the same ultra-structure, we have introduced the ``3-under-2'' principle leading to the addition of a new  layer  as described in Figure \ref{fig:insertionCycleAssume}.
The application of this principle leads to the structure of configuration \textbf{B} in Figure \ref{fig:structureAssume}, the configuration \textbf{A} on the same figure is not reachable by the chosen principle.

\begin{figure}[hbtp]
  \centering
  \includegraphics[width=5cm]{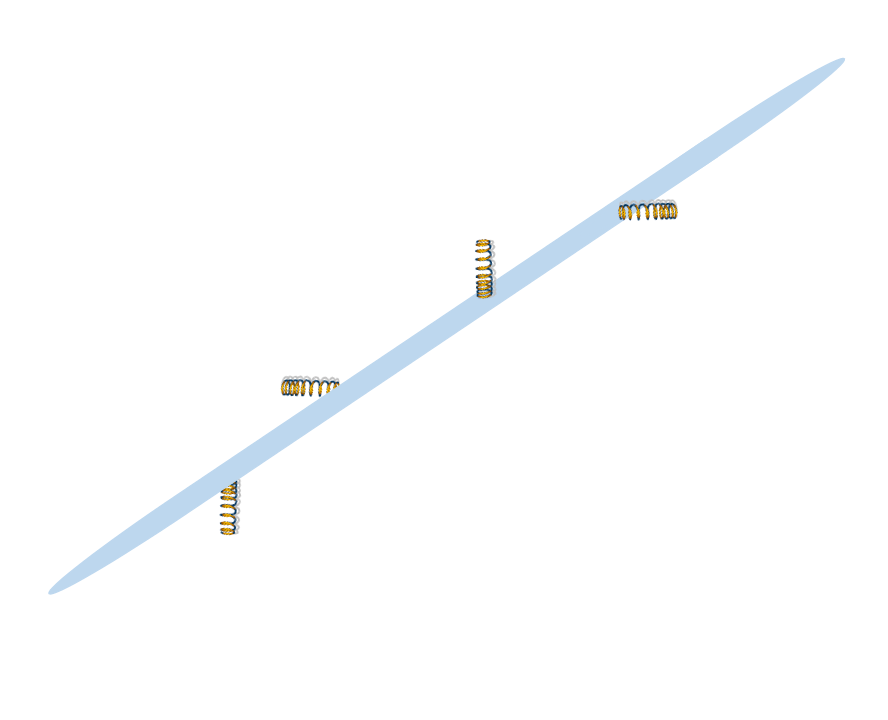}
  \caption{A realistic model of a PG strand}
  \label{fig:pgStrand}
\end{figure}

\begin{figure}[hbtp]
  \centering
  \includegraphics[width=13cm]{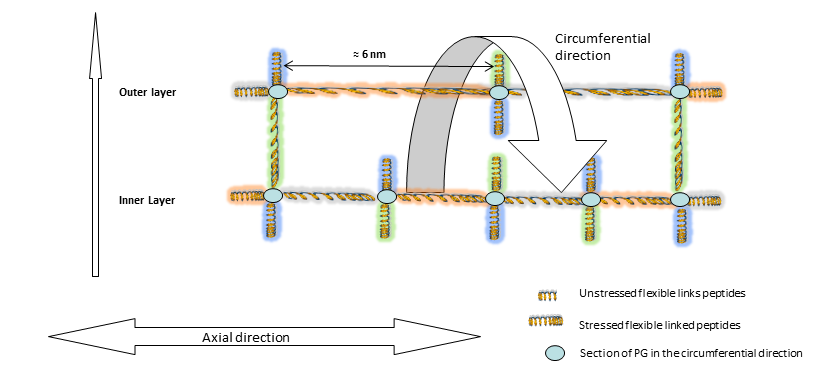}
  \caption{Hypothesized bi-layers CW: a colored aura represents a depth.}
  \label{fig:ourBiLayer}
\end{figure}

The previous derivations can be made with the usual model of the PG,  depicted in Figure \ref{fig:pgStrand}. Under this more realistic description, we obtain the ultra-structure depicted in Figure \ref{fig:ourBiLayer} and we will show by simulation that the ``3-under-2'' principle guarantees the regularity of the CW structure.

\section{Simulation}
For a mono-layer CW, all the pressure is transferred, through the membrane, to the only one existing layer which dissipates all the pressure energy. For a multi-layers CW, the situation is more complex as each layer dissipates only a fraction of the pressure energy. The problem is well summarized in the case for a bi-layers CW. The pressure is transferred to the innermost layer which bears a fraction of it via its internal structure; since both layers do not penetrate each other, the remaining part of the pressure is transferred to the outermost layer via an interaction of repulsive nature, this outermost layer bearing this remaining part. To simulate such features, there is a need for a dedicated and complex algorithm such as the ones used in clothes simulation where the clothes should not penetrate the person's body wearing them \cite{BaW:98}.

\begin{figure}[hbtp]
  \centering
  \includegraphics[width=13cm]{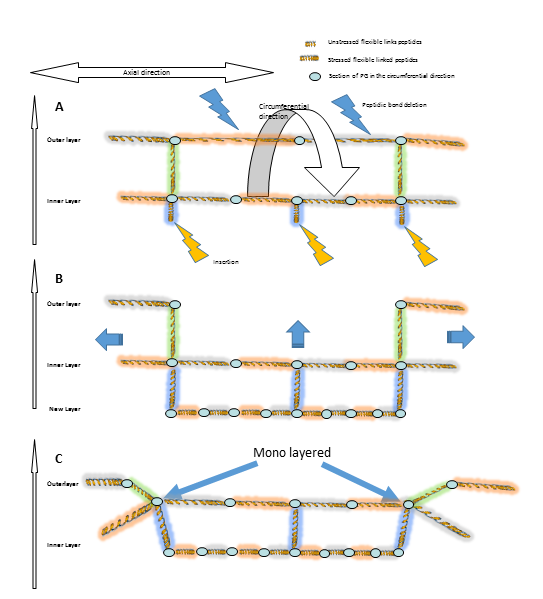}
  \caption{An elongation cycle showing two mono-layered strands (all the unnecessary potential peptidic bonds were deleted for clarity). (\textbf{A}) The original structure with the insertions and deletions. (\textbf{B}) The resulting CW ultra-structure: the newly PG strands and the ones above them are pushed up by the osmotic pressure while the CW elongates. (\textbf{C}) The extreme strands becomes mono-layered during the process even if a completed new layer would result in a bi-layers CW everywhere.}
  \label{fig:cwCycle}
\end{figure}

Nonetheless, our objective here is less demanding, we would like to validate the ``3-under-2'' principle which has been proposed only recently and not yet validated by simulation. Indeed, if for a complete layer the proposed model is coherent, the dynamic part of the CW elongation calls for a validation. In Figure \ref{fig:cwCycle}, it is shown that inserting a patch in a regular structure results in some PG strands to be mono-layered and that this strands are the only ones with all peptidic bonds actually connected to another PG.
This is a dynamical feature which disappears when the new layer has been completed as the regular nature of the CW structure will result in a two times longer structure with the same motif. Moreover the number of layer to layer peptidic bridges was shown above to be two times less than expected. This could lead to a potential fragility of the structure. Therefore, these layer to layer peptidic bridges may play an important role in the dynamical process of elongation.

\subsection{Simulation characteristics}
We propose a simple simulation on a regular structure in order to keep the simulation as simple as possible.

Since the circumference of the outermost strands is necessarily greater than the one of the innermost strands: they are more stressed in the radial direction and bear a greater fraction of the radial pressure. Due to their regularity, it is likely that each layer bears a constant fraction of this radial pressure. Moreover, the bacteria is assumed to regulate its osmotic pressure so that its value is constant in our simulations. The pressure has thus been split and applied on both layers separately saving the computation time of the interactions between the membrane and the innermost layer and between the layers themselves. From a computational point of view, the mono-layered strands belong to both layers so that they effectively bear all the pressure.

\begin{figure}[hbtp]
  \centering
  \includegraphics[width=7cm]{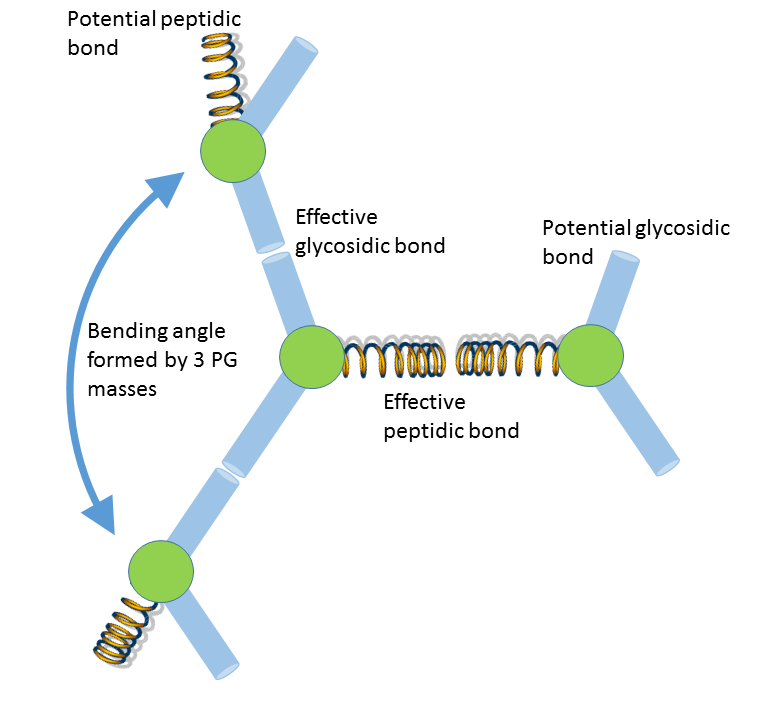}
  \caption{A PG is modelled as a mass (green) with two potential glycosidic bonds (blue) and one potential peptidic bond (spring). Whenever two potential glycosidic bonds match, they form an effective glycosidic one, likewise when two potential peptidic bonds match, they form an effective peptidic one. Each effective bond (either glycosidic or peptidic) stores an energy that creates a force applied on the two PG masses forming the bond. Whenever a PG has its two effective glycosidic bonds, a bending angle is created which stores an energy and creates a force applied on the three PG masses forming the angle.}
  \label{fig:pg}
\end{figure}

A PG is modelled as a mass that has potentially three bonds to other PG: 2 glycosidic bonds and 1 peptidic bridge (see Figure \ref{fig:pg}). The chain of glycosidic bonds forms a PG strand.

Moreover, as the insertion is not performed on the both poles, they are not simulated. We assume that the first and last strands of CW lie in a plane described by a mechanical frame which can translate and rotate with respect to a reference (inertial) frame. Axial pressure acts on the surface delimited by the PGs belonging to a pole.

The simulation is further detailed in Appendix \ref{sec:simul}.

\subsection{Swelling up of initial structure}
\begin{figure}[hbtp]
  \centering
  \includegraphics[width=13cm]{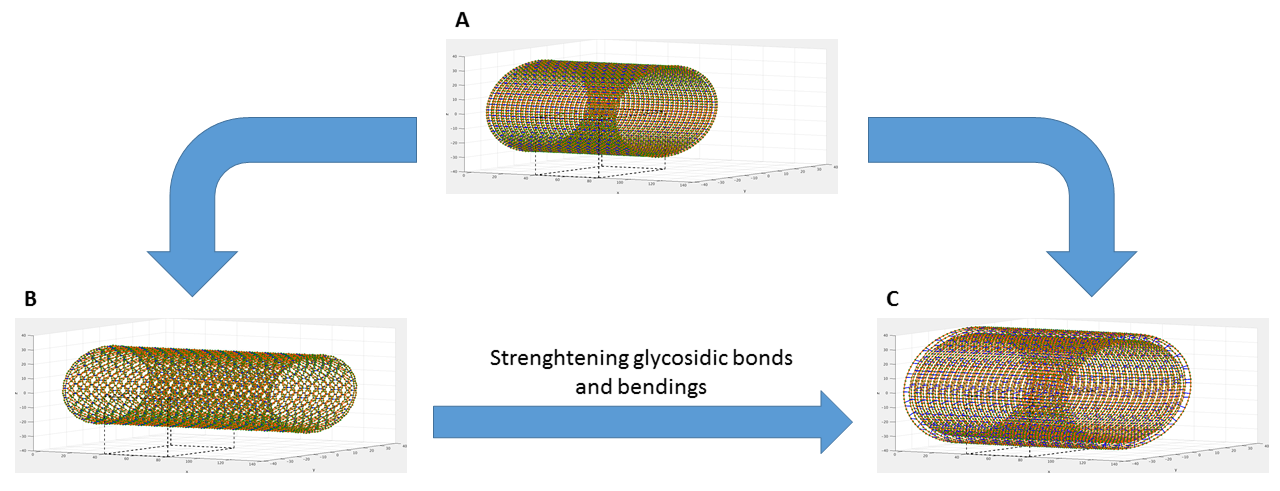}
  \caption{(\textbf{A}) Initial CW (PG mass in green, peptidic bonds in blue, glycosidic bonds in brown/red). (\textbf{B}) CW swelled up with the first parameter set. (\textbf{C}) CW swelled up with the second parameter set.}
  \label{fig:swellUp}
\end{figure}

\begin{figure}[hbtp]
  \centering
  \includegraphics[width=10cm]{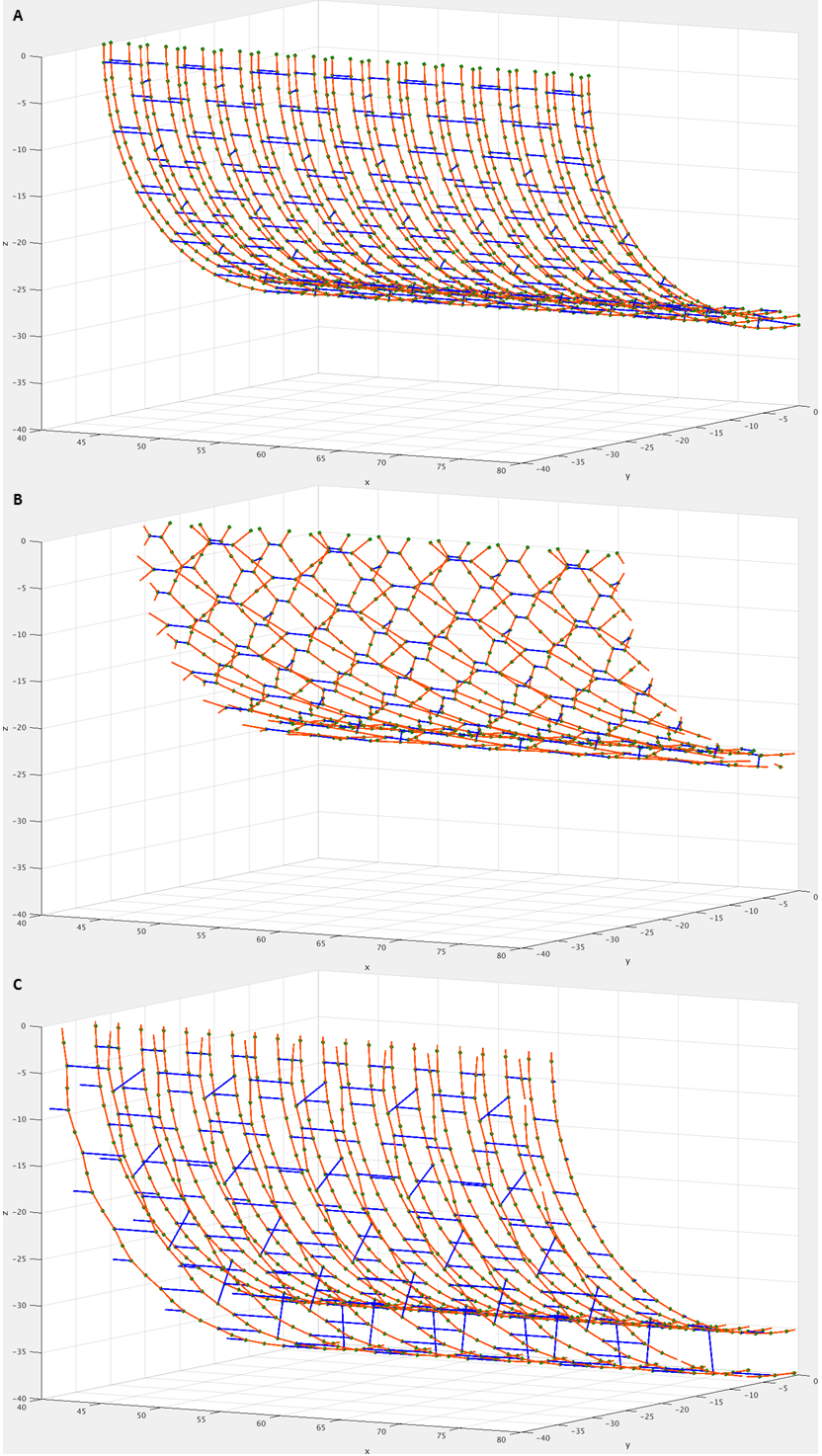}
  \caption{Zoom on the part of the space delimited by the dark dotted lines in Figure \ref{fig:swellUp}. (\textbf{A}) Initial CW (PG mass in green, peptidic bonds in blue, glycosidic bonds in brown/red). (\textbf{B}) CW swelled up with the first parameter set. (\textbf{C}) CW swelled up with the second parameter set.}
  \label{fig:swellUpZoom}
\end{figure}

We describe here the results of an initial swelling up (see Figure \ref{fig:swellUp} and Figure \ref{fig:swellUpZoom} for a zoom) with two sets of parameters values with the initial CW shown in panel \textbf{A}. The first set of values (see panel \textbf{B}) is the ones of \cite{HMWG+:08} and \cite{NGBJ:15}. The peptidic force was fitted with a polynomial of degree 9. We also used a pressure of 10 bars. In the second set of values (see panel \textbf{C}), we used a nonlinear force for the glycosidic bond equal to 5 times the one of the peptidic bond and increased the bending force by a factor 100.
As can be seen, for both sets, the CW ultra-structure is regular. With the second set, the bacteria is shorter and the hexagons are less pronounced due to a higher bending force, the overall structure being still regular. Note also that due to this higher bending force, the bacteria is larger since the circumference of a straighter strand is higher.

\subsection{Insertion of PG strands}
\begin{figure}[hbtp]
  \centering
  \includegraphics[width=14cm]{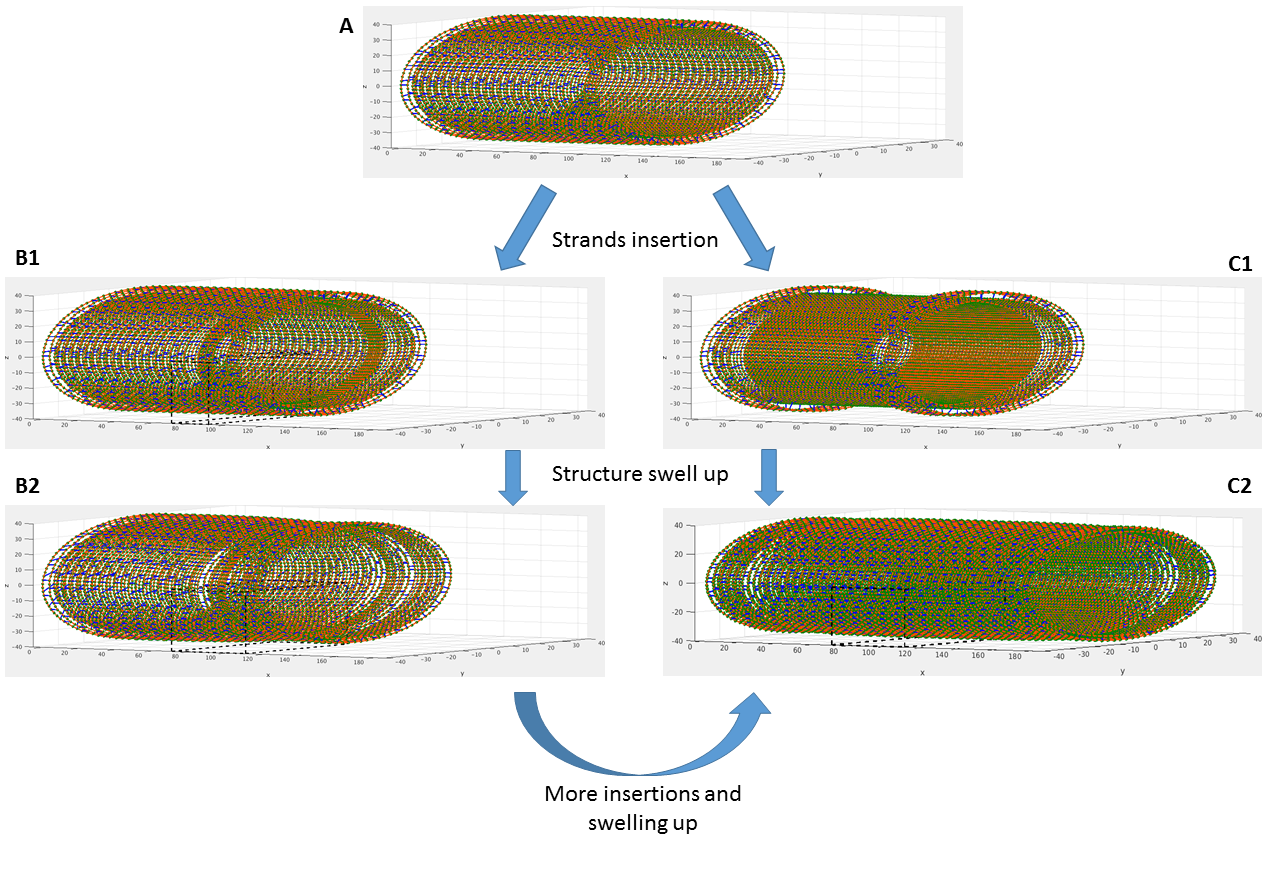}
  \caption{CW after strands insertion. (\textbf{A}) CW before insertion, from Figure \ref{fig:swellUp}, panel (\textbf{C}), rescaled for the need of this illustration. (\textbf{B1}) CW after a ``short''  insertion of 9 strands. (\textbf{B2}) CW after swell up of \textbf{B1}. (\textbf{C1}) CW after a ``long'' insertion of 81 strands. (\textbf{C2}) CW after swell up of \textbf{C1}.}
  \label{fig:insertPanel}
\end{figure}

From the results of Figure \ref{fig:swellUp}, panel (\textbf{C}), Figure \ref{fig:insertPanel} depicts the resulting CW structure after strands insertion.
In both panels \textbf{B2} and \textbf{C2}, the bacteria should be longer than in panel \textbf{A}. The length ratio can be approximated by the ratio of strands. The outermost layer of panel \textbf{A} is composed of 27 strands while it is composed of respectively 31 and 47 for respectively panel \textbf{B2} and \textbf{C2}, which gives a length ratio of around 1.1538 and 1.7692. With the length of the bacteria computed as the $x$ coordinate of the origin of the second pole, the obtained ratio are 1.1518 and 1.7427 which are consistent values.

\begin{figure}[hbtp]
  \centering
  \includegraphics[width=13cm]{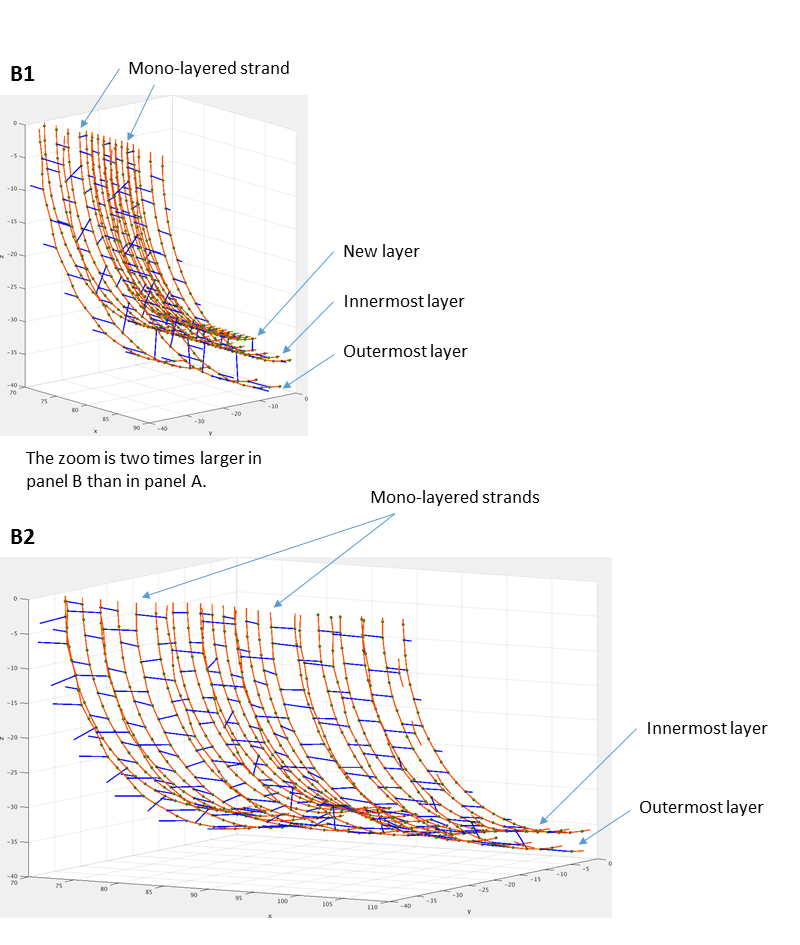}
  \caption{Zoom on the part of the space delimited by the dark dotted lines in Figure \ref{fig:insertPanel}. (\textbf{B1}) CW after insertion of 9 strands. (\textbf{B2}) CW after swell up of \textbf{B1}.}
  \label{fig:insertPanelZoom}
\end{figure}

\begin{figure}[hbtp]
  \centering
  \includegraphics[width=11cm]{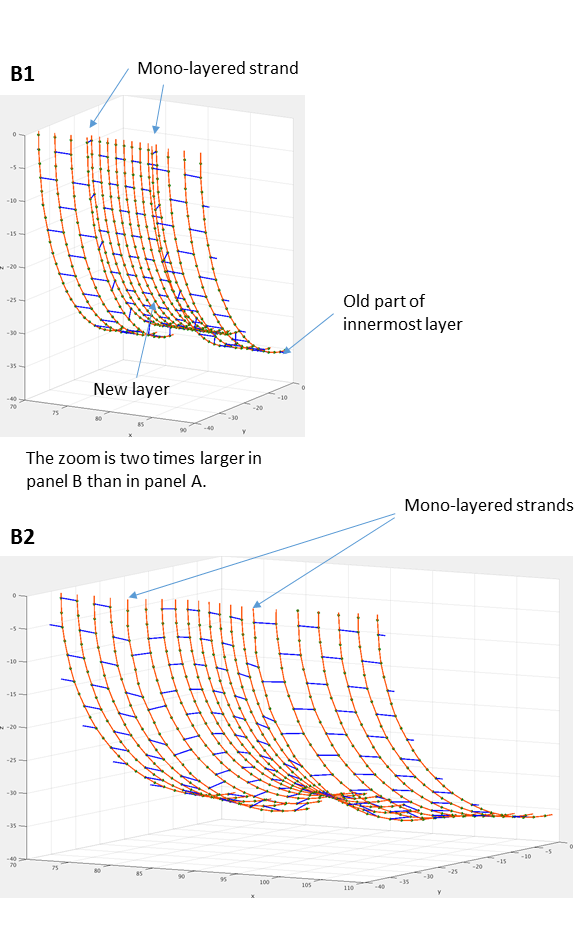}
  \caption{Zoom on the part of the space delimited by the dark dotted lines in Figure \ref{fig:insertPanel} for the innermost layer. (\textbf{B1}) CW after an insertion of 9 strands. (\textbf{B2}) CW after swell up of \textbf{B1}.}
  \label{fig:insertPanelOneLayerZoom}
\end{figure}

\begin{figure}[hbtp]
  \centering
  \includegraphics[width=13cm]{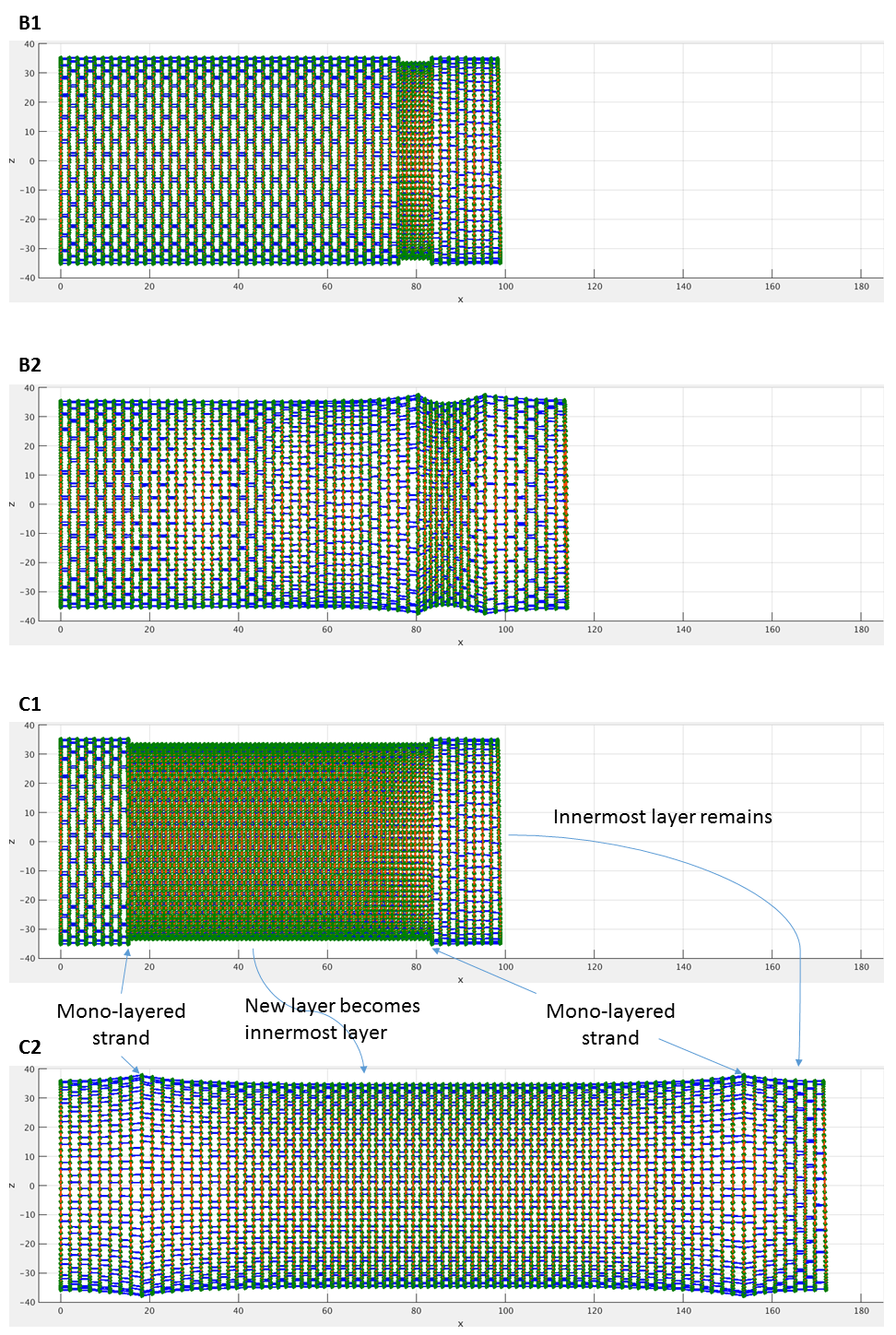}
  \caption{Innermost layer after insertion of 9 strands (panel \textbf{B1}) and swelling up (panel \textbf{B2}). Innermost layer after insertion of 81 strands (panel \textbf{C1}) and swelling up (panel \textbf{C2}).}
  \label{fig:longOneLayerInserted}
\end{figure}

Figure \ref{fig:insertPanelZoom} showing the insertion zone in the ``short'' insertion case confirms the displacement of the new layer as depicted in Figure \ref{fig:cwCycle} and the role of the layer to layer peptidic bridge; for clarity, Figure \ref{fig:insertPanelOneLayerZoom} displays the same figure but only for the innermost layer (likewise Figure \ref{fig:longOneLayerInserted} displays the innermost layer for both insertions showing at the same time the elongated CW).
The radius at the mono-layered strands are higher since they bear all the radial pressure alone. It can also be seen (especially in Figure \ref{fig:insertPanelOneLayerZoom}, panel \textbf{B2}) that these strands are less straight than the others since they also bear the axial pressure. Finally, two successive strands are closer together in the newly inserted part than in the old one. Indeed, when both layers are complete there are twice the number of peptidic bridges bearing the axial pressure in the innermost layer than in the outermost one. For the ``short'' insertion, inside the newly inserted part (see Figure \ref{fig:cwCycle}, panel \textbf{C}), there are 10 peptidic bridges bearing the axial pressure in the innermost layer for only 4 bridges in the outermost layer: the innermost bears, comparatively to complete layers, less of the axial pressure. The effect is less pronounced in the ``long'' insertion case as they are 82 peptidic bridges bearing the axial pressure in the innermost layer for 40 bridges in the outermost layer.

\begin{figure}[hbtp]
  \centering
  \includegraphics[width=13cm]{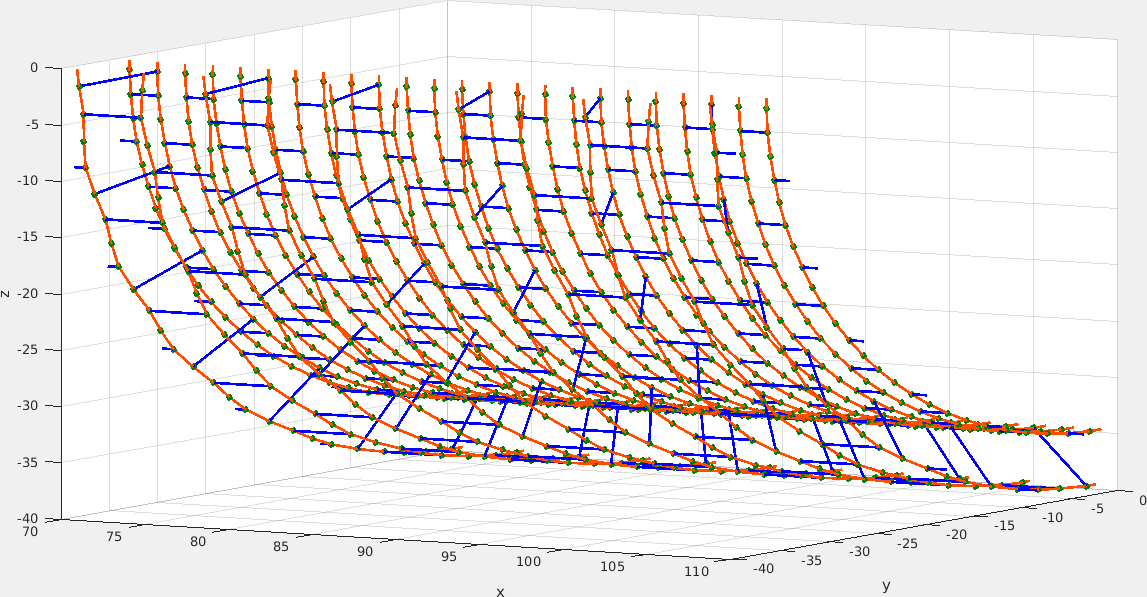}
  \caption{Zoom on the part of the space delimited by the dark dotted lines in Figure \ref{fig:insertPanel}, panel \textbf{C2}. After insertion cycles, the CW structure is the same as in the original one as shown in Figure \ref{fig:swellUpZoom}, panel \textbf{C}.}
  \label{fig:regStruct}
\end{figure}

Finally, we illustrate that the ``3-under-2'' principle leads to the regularity of the structure. Even if this can be figured out in Figure \ref{fig:insertPanelZoom}, it is clearer when more insertions are performed as depicted in Figure \ref{fig:regStruct}.

\bibliography{biblio-CW}
\bibliographystyle{plain}

\appendix

\section{Simulation Description} \label{sec:simul}
\subsection{Frames description}
We consider an absolute (inertial) frame ${\cal R}_0$ and two frames ${\cal R}_1$ and ${\cal R}_2$ linked to the poles. Each frame is described by an origin $o_r$ and three axes $\overrightarrow{i_r}$, $\overrightarrow{j_r}$ and $\overrightarrow{k_r}$, $r = 0,1,2$. The frames are orthonormal and direct. In the simulation, the frame ${\cal R}_1$ linked to the first pole is set and does not change. For convenience, this frame is equal to ${\cal R}_0$ and is thus also inertial. The second frame ${\cal R}_2$ can move and rotate with respect to ${\cal R}_0$. Let $p$ be a mass and its coordinates in ${\cal R}_0$ denoted by
$$
  p_{| {\cal R}_0} =
  \begin{bmatrix}
    x_{p | {\cal R}_0} \\
    y_{p | {\cal R}_0} \\
    z_{p | {\cal R}_0} \\
  \end{bmatrix}
  .
$$
Likewise, we denote
$$
  p_{| {\cal R}_1} =
  \begin{bmatrix}
    x_{p | {\cal R}_1} \\
    y_{p | {\cal R}_1} \\
    z_{p | {\cal R}_1} \\
  \end{bmatrix}
  \text{ and }
  p_{| {\cal R}_2} =
  \begin{bmatrix}
    x_{p | {\cal R}_2} \\
    y_{p | {\cal R}_2} \\
    z_{p | {\cal R}_2} \\
  \end{bmatrix}
$$
the coordinates of $p$ in ${\cal R}_1$ and ${\cal R}_2$. We have
$$
  p_{| {\cal R}_0} = p_{| {\cal R}_1}
  \text{ and }
  p_{| {\cal R}_0} = o_{2 | {\cal R}_0}
    + x_{p | {\cal R}_2} \overrightarrow{i_2}_{| {\cal R}_0}
    + y_{p | {\cal R}_2} \overrightarrow{j_2}_{| {\cal R}_0}
    + z_{p | {\cal R}_2} \overrightarrow{k_2}_{| {\cal R}_0}
$$
where $o_{2 | {\cal R}_0}$ is the coordinate of the origin of ${\cal R}_2$ in ${\cal R}_0$ and where $\overrightarrow{i_2}_{| {\cal R}_0}$, $\overrightarrow{j_2}_{| {\cal R}_0}$ and $\overrightarrow{k_2}_{| {\cal R}_0}$ are the coordinates of the axes of ${\cal R}_2$ in the frame ${\cal R}_0$. Considering a cell axis along $\overrightarrow{i_0}$, we have $x_{p | {\cal R}_1} = 0$ for a mass in the first pole; likewise we have $x_{p | {\cal R}_2} = 0$ for a mass in the second pole.

The movement of the frame ${\cal R}_2$ with respect to the frame ${\cal R}_0$ is governed by the equations of kinematics of classical mechanics and can be decomposed into a translation (of the origin) and a rotation (of the frame around its origin).

\subsection{Overview of the algorithm}
At the beginning of the simulation, we assume that the origin of the second frame is the center of mass of the peptidoglycans in the second pole and that the first pole is in the plane formed by $(o_0,\overrightarrow{j_0},\overrightarrow{k_0})$. 
The algorithm is: at iteration $t$,
\begin{enumerate}
  \item compute forces (see below for the forces computation) in the frame ${\cal R}_0$ denoted $\overrightarrow{F_p}_{| {\cal R}_0}(t)$:
    \begin{enumerate}
      \item compute all forces -- except the axial pressure force on the second pole: glycosidic and peptidic bonds, glycosidic bendings and radial pressure;
      \item add the above forces to obtain the resultant on all PGs; 
      \item set the $x$ component of force on all masses in the first pole to 0 (the first pole does not move);
      \item compute axial pressure force on the second pole (no computation on the first pole as it does not move);
      \item add this longitudinal pressure force (divided by the number of PG in the second pole);
    \end{enumerate}
  \item update coordinates in the frame ${\cal R}_0$:
    \begin{enumerate}
      \item for all PGs \emph{not} in the second pole, update the coordinates: $p_{| {\cal R}_0}(t+1) = p_{| {\cal R}_0}(t) + \alpha \overrightarrow{F_p}_{| {\cal R}_0}(t)$ where $\alpha$ is the algorithm step length;
      \item update the frame ${\cal R}_2$ and the coordinates of PGs in the second pole: 
      \begin{itemize}
        \item project all the forces into the frame ${\cal R}_2 (t)$, that is compute $F_{px2}(t) = \overrightarrow{F_p}_{| {\cal R}_0}(t) . \overrightarrow{i_2}_{| {\cal R}_0}(t)$, $F_{py2}(t) = \overrightarrow{F_p}_{| {\cal R}_0}(t) . \overrightarrow{j_2}_{| {\cal R}_0}(t)$ and $F_{pz2}(t) = \overrightarrow{F_p}_{| {\cal R}_0}(t) . \overrightarrow{k_2}_{| {\cal R}_0}(t)$;
        \item update the frame ${\cal R}_2$: 
        \begin{enumerate}
          \item update the origin: $o_{2 | {\cal R}_0}(t+1/2) = o_{2 | {\cal R}_0}(t) + \alpha \sum F_{pz2}(t) \overrightarrow{k_2}_{| {\cal R}_0}(t)$;
          \item update the axes:
          \begin{itemize}
            \item compute the moment vector: $\overrightarrow{\omega}_{| {\cal R}_0} (t) = \sum 
            \left(
            F_{pz2}(t) \overrightarrow{k_2}_{| {\cal R}_0}(t)
            \right)
            \wedge
            \left(
            F_{px2}(t) \overrightarrow{i_2}_{| {\cal R}_0}(t)
            +
            F_{py2}(t) \overrightarrow{j_2}_{| {\cal R}_0}(t)
            \right)
            $;
            \item update the axes: compute $\left( \overrightarrow{i_2}_{| {\cal R}_0}, \overrightarrow{j_2}_{| {\cal R}_0}, \overrightarrow{k_2}_{| {\cal R}_0}\right)(t+1)$ from a rotation of $\left(\overrightarrow{i_2}_{| {\cal R}_0}, \overrightarrow{j_2}_{| {\cal R}_0}, \overrightarrow{k_2}_{| {\cal R}_0}\right)(t)$ around $\alpha \overrightarrow{\omega}_{| {\cal R}_0} (t)$, that is a rotation around $\frac{\overrightarrow{\omega}_{| {\cal R}_0} (t)}{\| \overrightarrow{\omega}_{| {\cal R}_0} (t) \|}$ of angle $\alpha \| \overrightarrow{\omega}_{| {\cal R}_0} (t) \|$;
          \end{itemize}
        \end{enumerate}
        \item update the coordinates of all PGs in the second pole in ${\cal R}_2$:
          $$
            \begin{array}{ccl}
              x_{p | {\cal R}_2}(t+1/2) & = & 0 \\
              y_{p | {\cal R}_2}(t+1/2) & = & y_{p | {\cal R}_2}(t) + \alpha F_{py2}(t) \\
              z_{p | {\cal R}_2}(t+1/2) & = & z_{p | {\cal R}_2}(t) + \alpha F_{pz2}(t) \\
            \end{array}
          $$
        \item update the coordinates of all PGs in the second pole in ${\cal R}_0$: $p_{| {\cal R}_0}(t+1) = o_{2 | {\cal R}_0}(t+1/2)
    + x_{p | {\cal R}_2}(t+1/2) \overrightarrow{i_2}_{| {\cal R}_0}(t+1)
    + y_{p | {\cal R}_2}(t+1/2) \overrightarrow{j_2}_{| {\cal R}_0}(t+1)
    + z_{p | {\cal R}_2}(t+1/2) \overrightarrow{k_2}_{| {\cal R}_0}(t+1)$;
        \item finish update: compute $o_{2 | {\cal R}_0}(t+1)$ as the center of mass of $p_{| {\cal R}_0}(t+1)$ of all PGs in the second pole 
            and update their coordinates in ${\cal R}_2(t+1)$ to obtain $y_{p | {\cal R}_2}(t+1)$ and $z_{p | {\cal R}_2}(t+1)$ and set $x_{p | {\cal R}_2}(t+1) = 0$, that is compute $y_{p | {\cal R}_2}(t+1)$ and $z_{p | {\cal R}_2}(t+1)$ such that $(y_{p | {\cal R}_2}(t+1)-y_{p | {\cal R}_2}(t+1/2))\overrightarrow{j_2}_{| {\cal R}_0}(t+1) + (z_{p | {\cal R}_2}(t+1) - z_{p | {\cal R}_2}(t+1/2)) \overrightarrow{k_2}_{| {\cal R}_0}(t+1) = o_{2 | {\cal R}_0}(t+1) - o_{2 | {\cal R}_0}(t+1/2)$.
      \end{itemize}
    \end{enumerate}
\end{enumerate}

\subsection{Forces computation}
The following derivations of forces are very similar to the ones found in \cite{HMWG+:08,NGBJ:15}.

\subsubsection{Useful formulae}
Let A and B be two masses, then
$$
  \frac{\partial \| \overrightarrow{AB} \|}{\partial A}_{| {\cal R}_0}
  \overset{\triangle}{=}
  \begin{bmatrix}
    \frac{\partial \| \overrightarrow{AB} \|}{\partial x_{A| {\cal R}_0}} \\[2ex]
    \frac{\partial \| \overrightarrow{AB} \|}{\partial y_{A| {\cal R}_0}} \\[2ex]
    \frac{\partial \| \overrightarrow{AB} \|}{\partial z_{A| {\cal R}_0}} \\[2ex]
  \end{bmatrix}
  =
  - \frac{\overrightarrow{AB}_{| {\cal R}_0}}{\| \overrightarrow{AB} \|}
$$
with $\overset{\triangle}{=}$ meaning equal by definiton and
$$
  \frac{\partial \| \overrightarrow{AB} \|}{\partial B}_{| {\cal R}_0}
  =
  \frac{\overrightarrow{AB}_{| {\cal R}_0}}{\| \overrightarrow{AB} \|}
  =
  - \frac{\partial \| \overrightarrow{AB} \|}{\partial A}_{| {\cal R}_0}
  .
$$

\subsection{Bond}
We consider two masses (A and B) linked by a spring, that is A and B are linked by either a glycosidic or a peptidic bond) whose energy is given by
$$
  E = \frac{1}{2} \sum_{i=2}^n k_i \left( \| \overrightarrow{AB} \| - l_0 \right)^i
$$
where $\| \overrightarrow{AB} \|$ is the euclidean distance between A and B (independent of the chosen frame) and where $l_0$ is the spring length at rest.

Then the force applied by the spring stiffness on A is given by
$$
  \overrightarrow{F_A}_{| {\cal R}_0}
  =
  - \frac{\partial E}{\partial A}_{| {\cal R}_0}
  =
  - \frac{1}{2} \sum_{i=2}^n ik_i \left( \| \overrightarrow{AB} \| -l_0 \right)^{i-1} \frac{\partial \| \overrightarrow{AB} \|}{\partial A}_{| {\cal R}_0}
$$
leading to
$$
\boxed{
  \overrightarrow{F_A}_{| {\cal R}_0}
  =
  \left( \sum_{i=2}^n \frac{ik_i}{2} \left( \| \overrightarrow{AB} \| -l_0 \right)^{i-1} \right)
  \frac{\overrightarrow{AB}_{| {\cal R}_0}}{\| \overrightarrow{AB} \|}
  .
  }
$$
Likewise, the force applied by the spring stiffness on B is given by
$$
\boxed{
  \overrightarrow{F_B}_{| {\cal R}_0}
  =
  - \left( \sum_{i=2}^n \frac{ik_i}{2} \left( \| \overrightarrow{AB} \| -l_0 \right)^{i-1} \right)
  \frac{\overrightarrow{AB}_{| {\cal R}_0}}{\| \overrightarrow{AB} \|}
  =
  - \overrightarrow{F_A}_{| {\cal R}_0}
  .
  }
$$

\subsubsection{Bending}
We consider three masses (A, B and C, with C in between of A and B) with C and A having a glycosidic bond as well as C and B. The bending energy is given by
$$
  E = \frac{1}{2} \sum_{i=2}^n k_i \left( \theta - \theta_0 \right)^i
$$
where
$$
  \theta
  =
  \widehat{\overrightarrow{CA},\overrightarrow{CB}}
  =
  acos \left(
  \frac{\overrightarrow{CA}}{\| \overrightarrow{CA} \|}
  .
  \frac{\overrightarrow{CB}}{\| \overrightarrow{CB} \|}
  \right)
$$
is the angle formed by the three masses (independent of the chosen frame) and where $\theta_0$ is the angle at rest.

Then the force applied by the bending on A is given by
$$
  \overrightarrow{F_A}_{| {\cal R}_0}
  =
  - \frac{\partial E}{\partial A}_{| {\cal R}_0}
  =
  - \frac{1}{2} \sum_{i=2}^n ik_i \left( \theta -l_0 \right)^{i-1} \frac{\partial \theta}{\partial A}_{| {\cal R}_0}
$$
Since
$$
  \frac{\partial cos(\theta)}{\partial A}_{| {\cal R}_0} = -sin(\theta)\frac{\partial \theta}{\partial A}_{| {\cal R}_0}
  ,
$$
$$
  \overrightarrow{F_A}_{| {\cal R}_0}
  =
  \left( \sum_{i=2}^n \frac{ik_i}{2sin(\theta)} \left( \theta -l_0 \right)^{i-1} \right) \frac{\partial cos(\theta)}{\partial A}_{| {\cal R}_0}
$$
And since
$$
  \frac{\partial cos(\theta)}{\partial A}_{| {\cal R}_0}
  =
  \frac{\overrightarrow{CB}_{| {\cal R}_0}}{\| \overrightarrow{CA} \| \| \overrightarrow{CB} \|}
  +
  \frac{\overrightarrow{CA} . \overrightarrow{CB}}{\| \overrightarrow{CB} \|} \left( -\| \overrightarrow{CA} \|^{-2} \frac{\overrightarrow{CA}_{| {\cal R}_0}}{\| \overrightarrow{CA} \|} \right)
$$
leading to
$$
\boxed{
  \overrightarrow{F_A}_{| {\cal R}_0}
  =
  \left( \sum_{i=2}^n \frac{ik_i}{2sin(\theta)} \left( \theta -l_0 \right)^{i-1} \right)
  \frac{1}{\| \overrightarrow{CA} \|}
  \left(
  \frac{\overrightarrow{CB}_{| {\cal R}_0}}{\| \overrightarrow{CB} \|}
  -
  cos(\theta) \frac{\overrightarrow{CA}_{| {\cal R}_0}}{\| \overrightarrow{CA} \|}
  \right)
  .
  }
$$
Likewise, the force applied by the bending on B is given by
$$
\boxed{
  \overrightarrow{F_B}_{| {\cal R}_0}
  =
  \left( \sum_{i=2}^n \frac{ik_i}{2sin(\theta)} \left( \theta -l_0 \right)^{i-1} \right)
  \frac{1}{\| \overrightarrow{CB} \|}
  \left(
  \frac{\overrightarrow{CA}_{| {\cal R}_0}}{\| \overrightarrow{CA} \|}
  -
  cos(\theta) \frac{\overrightarrow{CB}_{| {\cal R}_0}}{\| \overrightarrow{CB} \|}
  \right)
  .
  }
$$
To obtain the force applied by the bending on C, we need
$$
  \frac{\partial cos(\theta)}{\partial C}_{| {\cal R}_0}
  =
  \frac{- \overrightarrow{CB} - \overrightarrow{CA}}{\| \overrightarrow{CA} \| \| \overrightarrow{CB} \|}
  -
  \frac{\overrightarrow{CA} . \overrightarrow{CB}}{(\| \overrightarrow{CA} \| \| \overrightarrow{CB} \|)^2}
  \left(
  - \| \overrightarrow{CB} \| \frac{\overrightarrow{CA}}{\| \overrightarrow{CA}\|}
  - \| \overrightarrow{CA} \| \frac{\overrightarrow{CB}}{\| \overrightarrow{CB}\|}
  \right)
$$
so that
$$
\boxed{
\begin{array}{ccl}
  \overrightarrow{F_C}_{| {\cal R}_0}
  & = &
  \left( \sum_{i=2}^n \frac{ik_i}{2sin(\theta)} \left( \theta -l_0 \right)^{i-1} \right) \times \dots \\
  & &
  \frac{1}{\| \overrightarrow{CA} \| \| \overrightarrow{CB} \|}
  \left(
  cos(\theta) \left( \| \overrightarrow{CB} \| \frac{\overrightarrow{CA}_{| {\cal R}_0}}{\| \overrightarrow{CA} \|} + \| \overrightarrow{CA} \| \frac{\overrightarrow{CB}_{| {\cal R}_0}}{\| \overrightarrow{CB} \|} \right)
  - \overrightarrow{CA} - \overrightarrow{CB}
  \right)
\end{array}
  .
  }
$$

\subsubsection{Pressure}
\begin{figure}[hbtp]
  \centering
  \includegraphics[width=10cm]{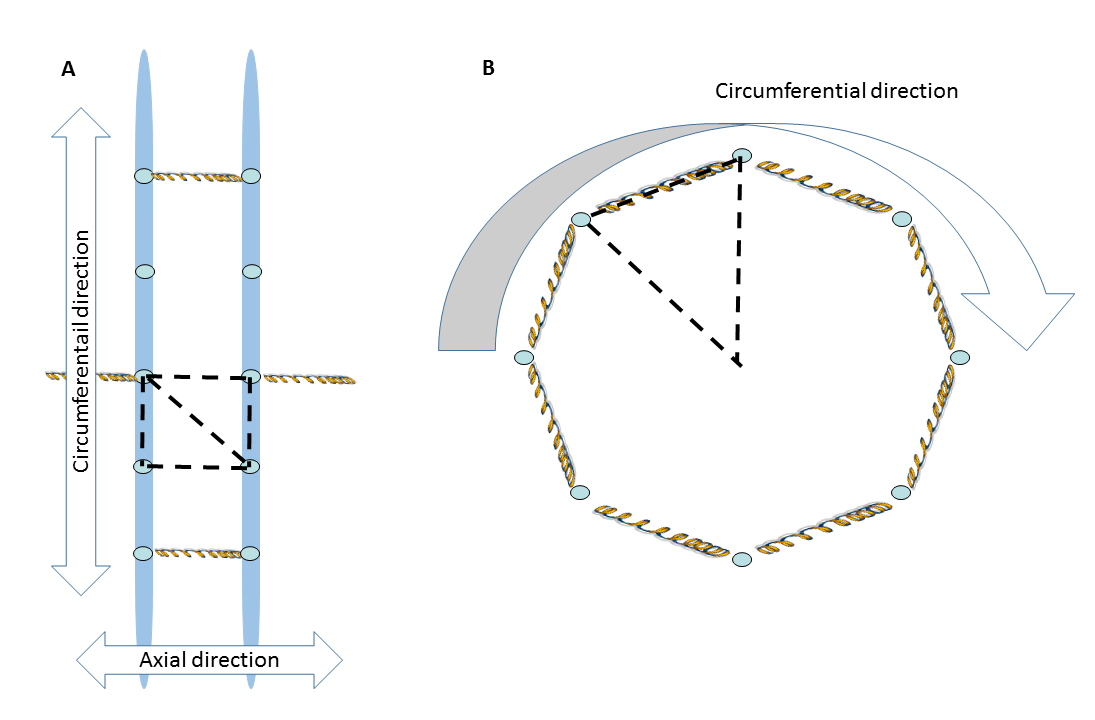}
  \caption{Surface for pressure computation. (\textbf{A}) Radial pressure. (\textbf{B}) Axial pressure.}
  \label{fig:cwTriangle}
\end{figure}

Let $P$ denote the (osmotic) pressure in the bacteria, more precisely the difference of pressure between the inside of the bacteria and its outside. Let $p_1$, $p_2$ and $p_3$ denote three points forming a surface of pressure. Then the pressure force is computed as
$$
  \overrightarrow{F}_{| {\cal R}_0} = \frac{P}{2} \overrightarrow{p_1p_2} \wedge \overrightarrow{p_1p_3}
  .
$$
The actual force applied on each PG depends on the actual number of physical masses forming the surface. The triangles chosen for the radial and axial pressure are plotted in Figure \ref{fig:cwTriangle} so that the actual number of PGs is 3 for the radial pressure. For the axial pressure, see the algorithm overview in Appendix A.2.

\end{document}